# Isolated Photonic Flatband with the Effective Magnetic Flux in A Synthetic Space including the Frequency Dimension

*Danying Yu[1], Luqi Yuan[1*], Xianfeng Chen[1,2]*

*[1]State Key Laboratory of Advanced Optical Communication Systems and Networks, School of Physics and Astronomy, Shanghai Jiao Tong University, Shanghai, 200240, China*

*[2]Collaborative Innovation Center of Light Manipulations and Applications, Shandong Normal University, Jinan, 250358, China*

[*yuanluqi@sjtu.edu.cn](mailto:yuanluqi@sjtu.edu.cn)

**Abstract:** The exploration of flatband in photonics is fundamentally important, aiming to control the localization of light for potential applications in optical communications. We study the flatband physics in a synthetic space including the frequency axis of light. A ring-resonator array is used to construct a synthetic Lieb-type lattice, where the modulation phase distribution supports a locally non-zero effective magnetic flux pointing into the synthetic space. We find that the flatband is isolated from other dispersive bands and the light can still be localized even with the perturbation from the group velocity dispersion of the waveguide. Our work points out a route towards manipulating the localization performance of a wavepacket of light along both the spatial dimension and the frequency dimension, which holds the potential implication for controlling the storage of optical information in dispersive materials.

# 1. Introduction

In photonics, it is of fundamental importance for the manipulation of light on demand. For example, tunable localization as well as transmission of light holds potential applications for optical communications and quantum information processing.[1,2] The use of the flatband in photonics has been shown to provide a powerful approach for localizing the wavepacket of light against small perturbations, due to the lack of dispersion of the bandstructure in the momentum space.[3,4] Moreover, gapless edge states is possible to be achieved in Lieb lattice by adding the effective magnetic field.[4] On the other hand, nontrivial flatbands in the condensed matter physics has been suggested to realize fractional quantum hall effect, which therefore attracts broad interest.[5-7] Many researches have been conducted to explore a variety of flatbands within different photonic structures, and seek the efficient control of light through complex photonic geometries,[8-11] non-Hermitian couplings,[12,13] the nontrivial topology,[6] and etc.[14-18] In particular, the isolation of the flatband gapped from other bands with dispersions is strongly desired for further robustness from disorders.[19]

The concept of synthetic dimensions in photonics arises and attracts a broad audience.[20,21] By making the analogy of physics in the real space, the optical phenomena in the synthetic space provide unique opportunities of manipulating not only the spatial propagation of light, but also the conversion of internal degrees of freedoms of light, such as the optical frequency and the orbital angular momentum.[20-26] Recently, the field of photonic synthetic dimensions has been pushed ahead thanks to several important experimental measurements including the first demonstration of photonic topological insulator in the synthetic space,[27] the manipulation of the spectra of light along the synthetic dimension,[28-30] and the observation of the effective magnetic flux in synthetic space.[31] As one of many candidates for synthesizing the extra dimension, the creation of synthetic dimension along the frequency axis of light in a ring resonator undergoing the dynamic modulation is a powerful approach.[23,24,30,31] In such designs, the arrangement of modulation phases provides the extra capability for generating the effective gauge field for photons on demand,[20] including the electric field,[32] the magnetic field,[22-24] and the local magnetic flux,[33] leading to the active control on the evolution of the wavepacket for photons spatially and spectrally, which are practically important for optical communications and information processing.[1,2]

In this paper, we propose to generate the isolated photonic flatband in a synthetic space by applying an effective magnetic flux for photons. The flatband in the synthetic space shows an efficient localization effect even with the intrinsic group-velocity dispersion (GVD) inside waveguide/fiber. The external dynamic modulation can actively control the propagation of wavepacket in synthetic space. In particular, we study a coupled-ring system, composed of two types of rings. By modulating one type of rings, a synthetic two-dimensional space including one spatial dimension and one frequency dimension is constructed, in which a Lieb-type[10] synthetic lattice has been created to exhibit a flatband. With modulation phases as $\pm\pi/4$ being distributed periodically along the spatial dimension in design, we introduce a locally non-zero effective magnetic flux in the synthetic space, which open gaps in the bandstructure and therefore isolate the flatbands. Such isolated flatband provides a unique opportunity for controlling the localization of light, and also is found to be robust against small waveguide's GVD. We simulate the propagation of the wavepacket in the synthetic space. Through active control the external modulation, we achieve the manipulation of the propagation of light. Our work paves the route for actively manipulating the efficient localization of light along both spatial and spectral axes in a photonic structure even with GVD inside the waveguide/fiber, which holds a promise for on-chip applications such as the storage of photons on demand.

## 2. Model without the group velocity dispersion

We start by considering a one-dimensional ring resonator array including two types of rings as shown in **Figure 1**a, labelled by $\alpha$ and $\beta$, respectively in each pair. Each ring is composed by the single-mode waveguide. We set the length of the ring $\alpha$ is twice long as the length of the ring $\beta$, i.e., $L_\alpha = 2L_\beta$, where $L_\alpha(L_\beta)$ is the circumference of the type $\alpha(\beta)$ ring. We assume that GVD of the waveguide is zero for the moment, and will take it into consideration in the later part of this paper. Hence the resonant frequency $\omega_{l,\alpha(\beta)}$ for the $l$th resonant mode in the type $\alpha(\beta)$ ring is:[23]

$$\omega_{l,\alpha(\beta)} = \omega_0 + l\Omega_{\alpha(\beta)} \qquad (1)$$

where $l$ is an integer, and $\Omega_{\alpha(\beta)} = 2\pi c/L_{\alpha(\beta)}n_{eff}$ is the free spectral range with $n_{eff}$ being the effective refractive index (which gives $2\Omega_\alpha = \Omega_\beta$). $\omega_0$ is the reference resonant frequency. In each type $\alpha$ ring, we place an electro-optic modulator performing the phase modulation with a time-dependent transmission coefficient:[33]

$$T = e^{i2\eta[\cos(\Omega t/2 + \phi_m)]} \qquad (2)$$

where $\eta$ is the modulation amplitude, $\Omega/2$ is the modulation frequency, and $\phi_m$ is the modulation phase [see Figure 1a]. We assume the modulation is resonant, i.e., $\Omega/2 = \Omega_\alpha$, which leads to the efficient exchange of the energy between resonant modes with the frequency difference $\Omega/2$, and hence couples adjacent resonant modes in each type $\alpha$ ring.[23] There is no modulator in each type $\beta$ ring so resonant modes inside each type $\beta$ ring remain discrete. The energies of light in two nearby rings are coupled through the evanescent wave for resonant modes at the same frequency. For simplicity, we assume that $\omega_0 = 0$ and use three types of modes ($A_{m,n}$, $B_{m,n}$, and $C_{m,n}$) to label all the resonant modes in the system. $A_{m,n}$ and $C_{m,n}$ corresponds to the $m$th resonant mode at the resonant frequency $\omega_{2n,\alpha}$ and $\omega_{2n+1,\alpha}$ respectively in the type $\alpha$ ring. $B_{m,n}$ corresponds to the $m$th resonant mode at the resonant frequency $\omega_{n,\beta}$ in the type $\beta$ ring [see Figure 1b for details]. Here $n$ is an integer. One can find that $\omega_{2n,\alpha} = \omega_{n,\beta} = n\Omega$ for modes $A_{m,n}$ and $B_{m,n}$, while $\omega_{2n+1,\alpha} = n\Omega + \Omega/2$ for modes $C_{m,n}$.

One can write the corresponding Hamiltonian for this coupled-ring system as:

$$H = \sum_{m,n}\{\omega_n(a_{m,n}^\dagger a_{m,n} + b_{m,n}^\dagger b_{m,n}) + (\omega_n + \Omega/2)c_{m,n}^\dagger c_{m,n}\} + \kappa\sum_{m,n}(a_{m,n}^\dagger b_{m,n} + a_{m,n}^\dagger b_{m-1,n} + h.c.) + 2g\cos(\Omega t/2 + \phi_m)\sum_{m,n}(a_{m,n}^\dagger c_{m,n} + a_{m,n}^\dagger c_{m,n-1} + h.c.) \quad (3)$$

where $a_{m,n}^\dagger$, $b_{m,n}^\dagger$, $c_{m,n}^\dagger$ ($a_{m,n}$, $b_{m,n}$, $c_{m,n}$) are the creation (annihilation) operators for the mode $A_{m,n}$, $B_{m,n}$, $C_{m,n}$ respectively. $\kappa$ is the coupling coefficient between modes having the same frequencies in two adjacent rings. $g = \eta\Omega/4\pi$ is the coupling coefficient between two nearest-neighbor resonant modes in the type $\alpha$ ring. By replacing $a_{m,n} \to a_{m,n}e^{i\omega_m t}$, $b_{m,n} \to b_{m,n}e^{i\omega_m t}$, $c_{m,n} \to c_{m,n}e^{i(\omega_m + \Omega/2)t}$, respectively, and taking the rotating-wave approximation under $g \ll \Omega$, one can rewrite the Hamiltonian (3) as:

$$\tilde{H} = \kappa\sum_{m,n}(a_{m,n}^\dagger b_{m,n} + a_{m,n}^\dagger b_{m-1,n}) + g\sum_{m,n}(a_{m,n}^\dagger c_{m,n}e^{-i\phi_m} + a_{m,n}^\dagger c_{m,n-1}e^{i\phi_m}) + h.c. \quad (4)$$

The Hamiltonian in Equation 4 describes a two-dimensional lattice structure in the synthetic space including the spatial and the frequency dimensions. Such synthetic lattice gives the Lieb-type lattice,[10] except for additional hopping phases $\phi_m$ in the couplings between two nearest-neighbor modes in each type $\alpha$ ring. The distribution of modulation phases therefore provides an important method for actively manipulating the light.

We explore the bandstructure from the Hamiltonian in Equation 4. With the assumption of both dimensions along the spatial and frequency axes being infinite, $k_x$ and $k_f$, which are the wave vectors reciprocal to spatial axis $x$ and frequency dimension $f$, respectively, are good quantum numbers. We consider the case with $\phi_m = 0$. The synthetic lattice has period of $d$ along the $x$ axis and $\Omega$ along the $f$ axis, where $d/2$ is the physical spacing between two adjacent rings. Hence there are three modes ($A_{m,n}$, $B_{m,n}$, $C_{m,n}$) in each unit cell. We can therefore write the Hamiltonian in Equation 4 into the $k$-space:

$$\widetilde{H}_{k_x,k_f} = 2\kappa a^\dagger_{k_x,k_f} b_{k_x,k_f} \cos\left(\frac{1}{2}k_x d\right) + 2g a^\dagger_{k_x,k_f} c_{k_x,k_f} \cos\left(\frac{1}{2}k_f \Omega\right)$$
$$+ 2\kappa b^\dagger_{k_x,k_f} a_{k_x,k_f} \cos\left(\frac{1}{2}k_x d\right) + 2g c^\dagger_{k_x,k_f} a_{k_x,k_f} \cos\left(\frac{1}{2}k_f \Omega\right) \tag{5}$$

We choose $\kappa = g$, and plot the bandstructure in the first Brillouin zone with $k_x \in [0, 2\pi/d]$ and $k_f \in [0, 2\pi/\Omega]$ in **Figure 2a**. One can see three bands, where the top and bottom ones are dispersive and the middle one is flat in the $k$-space. If the wavepacket of light is excited on the flatband, it is lack of diffusion along the spatial dimension and lack of dispersion along the frequency axis. Moreover, there is a degenerate point at $(k_x, k_f) = (\pi/d, \pi/\Omega)$, indicating that the energy in the flat band can be scattered through the degenerate point to one of the dispersive bands. For photonic systems operating at optical wavelengths, light usually exhibits a non-zero bandwidth. This means that, if the center frequency of light is chosen to excite the flatband, other frequency components within the bandwidth correspond to dispersive bands, which still leads to a portion of light propagating in the system. Hence the existence of the degenerate point prohibits the efficient localization of the light towards the potential implication for the storage of the optical information. The similar bandstructure with flatband has been reported in the Lieb lattice model in real space.[10]

To overcome the limitation described in the previous paragraph, we introduce the distribution of the modulation phase $\phi_m$ to open the degenerate point. The flatband is unchanged but gaps are open between the flatband and the bands with dispersion. In details, we set $\phi_m = -(+)\pi/4$ for $m$ being the odd (even) number. A nonzero magnetic flux is therefore created in the synthetic space including the frequency axis of light [see Figure 1c]. With such phase distribution, there are six modes in each unit cell. Hence the Hamiltonian in the $k$-space is:

$$\widetilde{H}_{k_x,k_f} = \kappa a^\dagger_{k_x,k_f,+} b_{k_x,k_f,+} e^{-ik_xd/2} + \kappa a^\dagger_{k_x,k_f,+} b_{k_x,k_f,-} e^{ik_xd/2} + \kappa a^\dagger_{k_x,k_f,-} b_{k_x,k_f,-} e^{-ik_xd/2}$$
$$+\kappa a^\dagger_{k_x,k_f,-} b_{k_x,k_f,+} e^{ik_xd/2} + 2g a^\dagger_{k_x,k_f,+} c_{k_x,k_f,+} \cos\left(\frac{1}{2}k_f\Omega + \frac{\pi}{4}\right) \quad (6)$$
$$+2g a^\dagger_{k_x,k_f,-} c_{k_x,k_f,-} \cos\left(\frac{1}{2}k_f\Omega - \frac{\pi}{4}\right)$$

Here +/- in operators refers to the odd/even $m$ in the synthetic lattice. We set all the other parameters the same as those in Figure 2a and plot the bandstructure from Equation 6 in Figure 2b. The Brillouin zone involves $k_x \in [0, \pi/d]$. There are six bands, four of which are dispersive in the synthetic $k$-space. The other two bands, separated from the four dispersive bands, are flat and also degenerate over the whole $k$-space. The gaps between dispersive bands and flatbands are $\sim g$, indicating that the localized state associated to the excited flatband is hard to be scattered to states at the dispersive band for small perturbations with scattering coefficient smaller than $0.5g$. Such isolated flatband in the synthetic space provides a unique opportunity for the efficient localization of the light on demand by actively turning on the modulations.

## 3. Model with the group velocity dispersion

Group velocity dispersion (GVD) is intrinsic in waveguide/fiber-based system. GVD can be used to set a boundary in the synthetic lattice,[23] but it also brings the effective harmonic potential to the system.[24] In a Lieb-type lattice, the introduction of such effective harmonic potential along the synthetic dimension may significantly change the flatness of the flatband, and then influence localization effect in both dimensions. We take the GVD of the waveguide into considerations, which we set to be zero previously. Although the GVD has been ignored in many previous researches,[23-25,32,33] the physics in the synthetic space including the frequency axis of light with the GVD is subtle and deserves the careful study. Furthermore, for the localization of the light by using the flatband without the dispersion in the momentum space, it is reasonable to explore physical consequences associated to the flatband in the system including the GVD of the waveguide. By including GVDs in each ring, one can rewrite Equation 4 as:

$$\widetilde{H}_D = \kappa \sum_{m,n} \left( a_{m,n}^\dagger b_{m,n} + a_{m,n}^\dagger b_{m-1,n} \right) + g \sum_{m,n} \left( a_{m,n}^\dagger c_{m,n} e^{-i\phi_m} + a_{m,n}^\dagger c_{m,n-1} e^{i\phi_m} \right) + h.c.$$
$$+ \sum_{m,n} \frac{D}{2}(n - n_0)^2 \left( a_{m,n}^\dagger a_{m,n} + b_{m,n}^\dagger b_{m,n} \right) + \sum_{m,n} \frac{D}{2}(n + 0.5 - n_0)^2 \left( c_{m,n}^\dagger c_{m,n} \right) \quad (7)$$

where $D$ refers to the contribution of the GVD of the waveguide, which gives the small detuning dependent on $n$ for modes near the reference $n_0$ mode at $\omega_{n_0}$.[24] We assume that GVDs in both rings are the same. For the case that two rings are composed by different waveguide, which may lead to different GVDs, is beyond the scope of this work but can be studied in the future.

From Equation 7, one notices that resonant modes in each ring are no longer equally spaced along the frequency axis of light. $k_f$ therefore is no longer a good quantum number. However, since we assume that GVDs in different rings are the same, the modes $A_{m,n}$ and $B_{m,n}$ in two nearest-neighbor rings still have same resonant frequencies, which makes the system still has a good periodicity along the $x$ direction and $k_x$ is a good quantum number. We calculate the projected bandstructure for a synthetic strip, which is infinite in the $x$ direction, but in the frequency dimension, $n \in [1,20]$ for modes $B$ and $C$ and $n \in [1,21]$ for modes $A$ with $n_0 = 11$. We set $\kappa = g$ and choose the phase $\phi_m$ being zero and $\phi_m = \pm \pi/4$, respectively, which corresponds to same parameters as those in Figure 2 with/without the effective magnetic flux. We introduce the GVD and use $D/g$=0, 0.01, 0.03, respectively to calculate projected bandstructures as shown in **Figure 3**. In Figure 3a-c, we set $\phi_m = 0$, so there is no magnetic flux. In bandstructures, one can see the middle band being flat and also a degenerate point at $\varepsilon/g = 0$ for the case of $D/g$=0 in Figure 3a, which is consistent with the bandstructure in Figure 2a. With the increase of the GVD, i.e., $D/g$=0.01 and 0.03 in Figure 3b and 3c, the projected middle band near $\varepsilon/g = 0$ becomes broader and the degenerate point disappear. This phenomenon indicates that not only the middle band loses its feature of exact flatness and also the excited state associated to the middle band becomes easier to be scattered to dispersive bands, which causes the localization being unstable while light evolves. As for the comparison, we plot the projected bandsturctures with the effective magnetic flux in Figure 3d-f. When $D/g$=0, the projected bandstructure is consistent with the bandstructure in Figure 2b, except for pairs of edge modes corresponding to boundaries at the synthetic frequency axis of the strip. The edge modes do not touch the flatband in the middle at $\varepsilon/g = 0$. For the case of $D/g$=0.01 as shown in Figure 3e, although the middle band becomes broader, it is still isolated from either the dispersive band or the edge mode. Comparing both cases of $D/g$=0.01 with/without the magnetic flux in the Figures 3b and 3e, one expects to see the effect of localization may still preserve in the case with the magnetic flux. With further increasing $D/g$ to 0.03 in Figure 3f, the gap between the middle band and the dispersive bands disappears, which is similar to the case of $D/g$=0.03 without the magnetic flux.

In addition, we calculate the bandstructure and project it along along the *x* axis while changing the GVD parameter *D* for the case with the effective magnetic flux in the synthetic space in **Figure 4**. The color corresponds to the densities of states (DOS) in photonics at each eigen-energy $\varepsilon$, which is normalized. At *D*=0, one can see a single point at $\varepsilon/g = 0$ corresponding to the maximum DOS. Moreover, two gaps with the width $\Delta\varepsilon/g \sim 1$ exist above and below this point. This feature is consistent with the bandstructure in Figure 3d. With the increase of *D*, gaps start to close and the bandstructure eventually becomes gapless near $D/g$~0.018. The phase transition from the gapped bandstructure to the gapless bandstructure has been shown. Nevertheless, one notices that even in the condition of the existence of small GVD, the existence of the effective magnetic flux keeps the gap open and isolate the middle band. Moreover, the corresponding DOS at the middle band near $\varepsilon/g = 0$ holds a large number for small GVD.

## 4. Numerical simulations to show the localization effect

Here we perform the numerical simulation to demonstrate the effect of the localization by the flatband under the effective magnetic flux in the synthetic space. We consider a system consisting of 21 large type α rings and 20 small type β rings. Moreover, in each type α ring, we take 41 resonant modes into the consideration, while in each type β ring, we take 21 resonant modes. Therefore, in the synthetic space, we have mode $A_{m,n}$ ($m \in [1,21]$, $n \in [1,21]$), $B_{m,n}$ ($m \in [1,20]$, $n \in [1,21]$), and $C_{m,n}$ ($m \in [1,21]$, $n \in [1,20]$), repectively. The reference mode is chosen at $n_0 = 11$. We excite the synthetic lattice by coupling the light through an external waveguide into the central ring (the 11th type α ring) at the frequency $\omega_{n_0} = 11\Omega$, i.e., to excite the single mode $A_{11,11}$ at the resonance.[23,33]

We consider the wave function of light:

$$|\varphi(t)\rangle = \sum_{m,n} [v_{a,m,n}(t)a^\dagger_{m,n} + v_{b,m,n}(t)b^\dagger_{m,n} + v_{c,m,n}(t)c^\dagger_{m,n}]|0\rangle \tag{8}$$

where $v_{a,m,n}$, $v_{b,m,n}$, $v_{c,m,n}$ is the amplitude of the photon state at the mode $A_{m,n}$, $B_{m,n}$, $C_{m,n}$, respectively. By applying the Schrödinger Equation $\frac{d}{dt}|\varphi\rangle = -i\widetilde{H}_D|\varphi\rangle$ with $\widetilde{H}_D$ defined in Equation 7, the intrinsic loss $\gamma$, and also the source $s$ excited at the mode $A_{11,11}$, one obtains the coupled-mode equations:

$$\begin{aligned}
\dot{v}_{a,m,n} &= -\gamma v_{a,m,n} - i\kappa(v_{b,m,n} + v_{b,m-1,n}) - ig(v_{c,m,n}e^{-i\phi_m} + v_{c,m,n-1}e^{i\phi_m}) \\
&\quad - iv_{a,m,n}\frac{D}{2}(n-n_0)^2 + s\delta_{m,11}\delta_{n,11} \\
\dot{v}_{b,m,n} &= -\gamma v_{b,m,n} - i\kappa(v_{a,m,n} + v_{a,m+1,n}) - iv_{b,m,n}\frac{D}{2}(n-n_0)^2 \\
\dot{v}_{c,m,n} &= -\gamma v_{c,m,n} - ig(v_{a,m,n}e^{i\phi_m} + v_{a,m,n+1}e^{-i\phi_m}) - iv_{c,m,n}\frac{D}{2}(n+0.5-n_0)^2
\end{aligned} \tag{9}$$

In simulations, we choose $\kappa = g$, $\gamma = 0.1g$, and $s = e^{i\Delta\varepsilon t}$, where $\Delta\varepsilon$ denotes the detuned frequency of the input light. The dissipation rate $\gamma$ is carefully chosen so the wavepacket of light can expand sufficiently in either the spatial or frequency dimension alone if the dispersive band is excited but the expansion does not reach to the boundary to avoid unwanted reflections. For the two-dimensional model as shown in Figure 1, steady-state simulations are performed and results are plotted in **Figure 5**. We first consider the case without the magnetic flux [see Figure 5a-c], which corresponds to bandstructures in Figure 3. For the case with $D=0$ and $\Delta\varepsilon = 0$, one sees the effect of the localization as shown in Figure 5a. The energy of light is stored in a few rings in the middle at the resonant modes near $n=11$. On the other hand, if we choose the excitation $\Delta\varepsilon/g = 0.3$ to excite the dispersive band in Figure 3a, the field spreads in the synthetic Lieb lattice in Figure 5b. Hence the excitation of the flatband requires the source field has single excitation frequency exactly at $\Delta\varepsilon = 0$. For a larger $D$ ($D/g$=0.03 and $\Delta\varepsilon = 0$) as shown in Figure 5c, one notices that the energy distribution of light propagates along the spatial (horizontal) dimension due to the fact that the flatband is destroyed by the effects of GVD. On the other hand, since GVD presents an effective harmonic potential along the frequency axis of light [see Equation 7], the conversion of resonant modes along the frequency (vertical) dimension is restricted.

On the other hand, in Figure 5d-f, we show the simulation results for similar cases but including the effective magnetic flux by preparing the phase distribution as shown in Figure 1c. When $D=0$ and $\Delta\varepsilon = 0$, one sees the localization effect of the energy for light in Figure 5d, which exhibits a better isolation behavior compared to the case without the magnetic flux in Figure 5a. Nevertheless, if we choose the excitation $\Delta\varepsilon/g = 0.3$ instead, since there is no band at this excitation frequency in the bandstructure as shown in Figure 3d, the flatband can still be excited and the energy for light is therefore localized as shown in Figure 5e. This localization effect preserves because of the bandgap open due to the effective magnetic flux in the synthetic Lieb lattice, which shows very different evolution patterns as compared to Figure 5b. Moreover, different evolution patterns are also found between Figure 5f and Figure 5c for the case with GVD at $D/g$=0.03 and $\Delta\varepsilon = 0$. As shown in Figure 5f, the photon state is restrained in a smaller area in the synthetic space, indicating that the effective magnetic flux can preserve the localization effect against small GVD along the frequency axis of light. In this case, even though the bandstructure in Figure 3f shows the close of the gap, the simulation shows that the most energy of light is still remaining in the center of the synthetic space. One can understand the phenomena from Figure 4, where the band near $\varepsilon/g \sim 0$ exhibits a large number of densities of states, which have been largely excited by the external source. By comparing simulation results in Figure 5, one can find that the effective magnetic flux indeed keeps the localization of energy for small GVD. As a note here, the dissipation rate $\gamma$ chosen in simulations makes us focus on the major energy distribution of the field with the time evolution and filter some small energy fluctuation of the field. For a significant large intensity of GVD, the intensive intersection between the flatband and the bands with the dispersion cannot be avoided, and hence the localization effect is no longer valid.

## 5. Numerical simulations to show the photon-storage effect

The localization effect in the synthetic space can be used to effectively control the propagation and the storage of light. To demonstrate that one can actively control the localization process and hence it leads to the possibility of the photon storage, we follow the same simulation procedures but set the zero dissipation for the simplicity. We choose the synthetic Lieb-type lattice with the effective magnetic flux. Additional synthetic dimension along the frequency axis of light can be turned on to generate the synthetic Lieb-type lattice to achieve localization, or be turned off for propagation along the spatial dimension.

In the simulation, the synthetic lattice is excited at the site $A_{1,11}$ by a pulsed source:

$$s = \left(\frac{1}{2} - \frac{1}{2}\tanh\left(\frac{1}{2}g(t-\tau)\right)\right)e^{i\Delta\varepsilon t} \qquad (10)$$

where $\tau$ denotes the pulse duration. We choose $\Delta\varepsilon = 0$ and $\Delta\varepsilon/g = 1.5$, respectively, and also $\tau = 6g^{-1}$, $D = 0$, $\gamma = 0$, with all the other parameters are the same as the simulation in Figure 5d. The simulation is performed until $t = 30g^{-1}$ and modulations inside rings are initially turned off, and are turned on during the time $t \in [10,20]g^{-1}$. Snapshots of the wavepacket evolution are provided at $t = 6g^{-1}$ for the pulse being off, $t = 10g^{-1}$ for modulations ready to be turned on, $t = 14g^{-1}, 20g^{-1}$ for modulations on, and $t = 24g^{-1}, 30g^{-1}$ for modulations being turned off, respectively.

**Figure 6** shows distributions of the wavepacket of light at different times for $\Delta\varepsilon = 0$ in Figure 6a-f and $\Delta\varepsilon/g = 1.5$ in Figure 6g-l, respectively. The middle site at the left boundary is excited in the synthetic lattice, which corresponds to the excitation at the resonant frequency $\omega_{n_0}$ in leftmost ring resonator. In Figure 6a-f for $\Delta\varepsilon = 0$, when no modulation at the time $t < 10g^{-1}$, there is no coupling along the frequency dimension and one can see the wavepacket is propagating towards the right at the frequency $\omega_{n_0}$. Interestingly, once the modulation is turned on to open the coupling channels between nearby resonant modes in type $\alpha$ rings, the propagation of the wavepacket stops and the energy of light is localized. The photon state at the flatband at $\Delta\varepsilon = 0$ is therefore prepared in the synthetic space. The major part of the energy does not spread along the frequency dimension. After $t = 20g^{-1}$, we turn off modulations and one sees that the major part of the wavepacket of light remains propagating towards the right. Our simulation is operated in the case with the existence of the effective magnetic flux, so a good localization effect exhibits. By actively turning on/off modulations, one is capable to see the effect of the photon storage.

As comparisons, in Figure 6g-l for $\Delta\varepsilon/g = 1.5$, the photon storage effect does not exist. When modulations are turned on during the time $t \in [10,20]g^{-1}$, the wavepacket of light spreads along the frequency axis, as the excitation at $\Delta\varepsilon/g = 1.5$ corresponds to the photon state associated to the dispersive band. Therefore, the wavepacket trapping, is sensitive to the input source, which is potentially useful for the selective photon storage.

## 6. Discussion and conclusion

Our proposal is valid for potential experiments on several optical platforms including the fiber-ring system[30,31] and on-chip silicon[34,35] or lithium niobite microrings.[36-39] For the fiber ring with the length ~10 m, commercial electro-optic modulators can be used with the modulation frequency at the order of 10 MHz. The modulation strength is tunable from 0 to ~1 MHz in the experiment.[30] For a weak modulation strength ~0.01Ω, the GVD in our consideration with $D/g \sim 0.01$ gives the number GVD $= -nD/c\Omega^2 \sim -10^3\ ps^2/m$.[40] On the other hand, for the on-chip design, the microring has the radius ~0.1-1mm, so the corresponding modulation is applied at a higher frequency at the order of 10 GHz. For weak modulation, it gives a GVD ~1 $ps^2/m$. In telecom band, both fiber and on-chip waveguide have smaller typical values of GVD, meaning that our design can indeed localize the light in realistic experimental platforms. Moreover, the good performance in localization with the large GVD holds a promise for applications with strongly dispersive materials with large GVD.[41,42]

We study a ring-resonator array that constructs the Lieb-type lattice in the synthetic space including the frequency axis of light. A flatband is created and hence the localization of light in both the spatial dimension and the spectral dimension is proposed. Moreover, by introducing the locally non-zero but totally zero effective magnetic flux in the synthetic lattice, we find an isolated flatband. Such isolated flatband shows a better localization effect in the synthetic space even if the ring's waveguide has significant group velocity dispersion, where the conventional localization effect from the Lieb structure is no longer valid. Turning on/off the external modulations can control the internal coupling between resonant frequency mode. Through actively controlling modulations along the frequency dimension, we can effectively manipulate the movement of the light. Our work opens a door for engineering isolated flatband in synthetic dimensions by actively manipulating the modulation phases, pointing to important implications for optical storage and optical communications with dispersive materials.

**Acknowledgements**

The author thanks Shanhui Fan and Avik Dutt for helpful discussions. This paper is supported by National Natural Science Foundation of China (11974245), National Key R&D Program of China (2018YFA0306301 and 2017YFA0303701), and Natural Science Foundation of Shanghai (19ZR1475700).

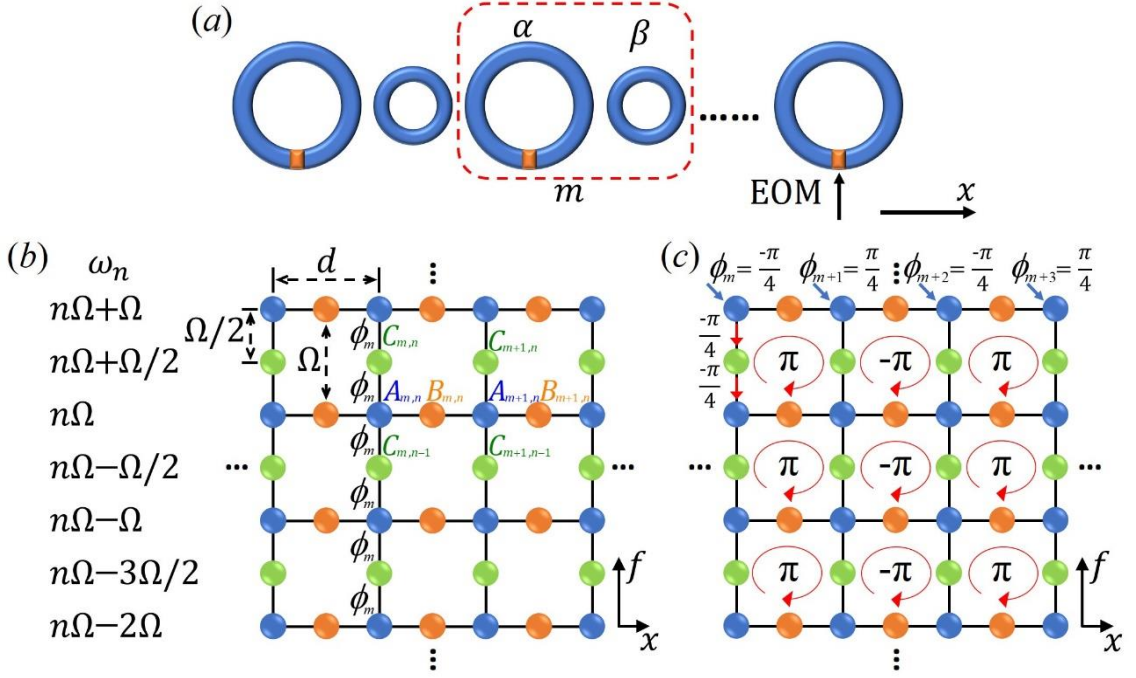

Figure 1. a) A ring resonator array consists of two type of rings, where large ring denotes the type $\alpha$ ring, and the small ring denotes the type $\beta$ ring. There are electro-optic modulators (EOMs) inside each type $\alpha$ ring. b) The array supports a synthetic lattice including one frequency dimension and one spatial dimension. c) By controlling the arrangement of modulation phases, the locally non-zero magnetic flux is constructed in the synthetic space.

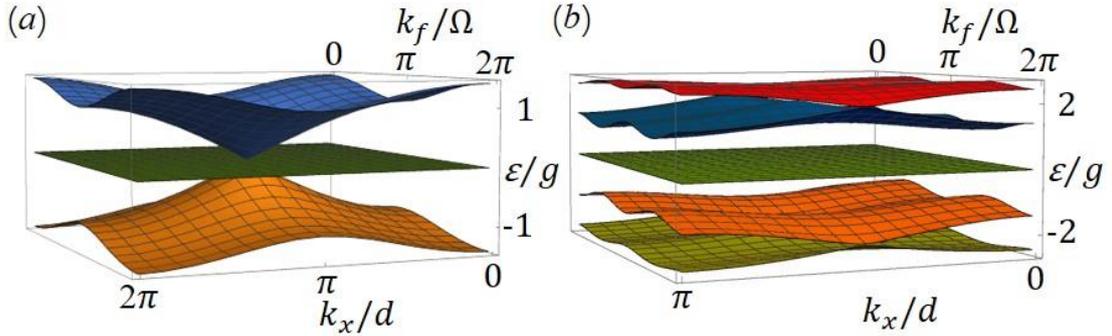

Figure 2. Bandstructures of the energy $\varepsilon$ as a function of momenta $k_x$ and $k_f$, which are reciprocal to the spatial and frequency dimensions, respectively. a) The bandstructure corresponds to the Lieb-type lattice in the synthetic space illustrated in Figure 1b. b) The bandstructure with the locally non-zero effective magnetic flux as shown in Figure 1c.

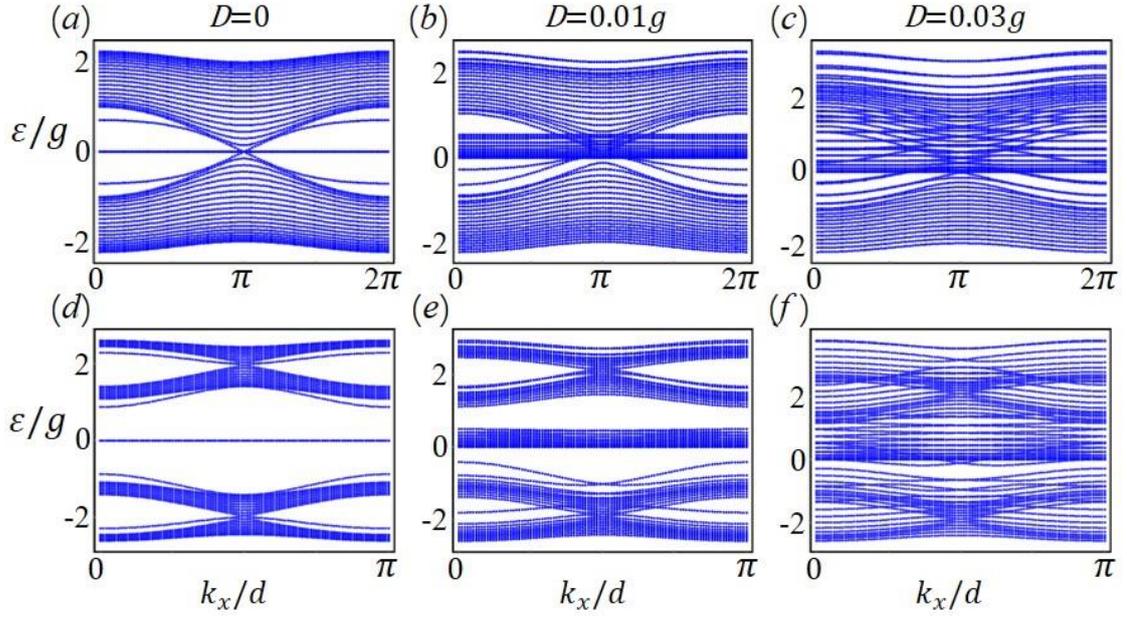

Figure 3. Projected bandstructure with the GVD of the waveguide included. a)-c) For the cases without the effective magnetic flux, the corresponding bandstructures with the GVD: $D/g$=0, 0.01, 0.03. d)-f) For the cases with the effective magnetic flux shown in Figure 1c, the corresponding bandstructures with the GVD: $D/g$=0, 0.01, 0.03.

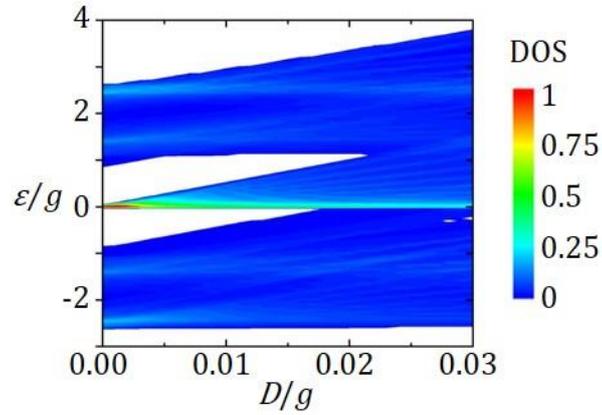

Figure 4. Projected bandstructure for the case with the effective magnetic flux with $D/g$ varying from 0 to 0.03. The color-encoded DOS represents the normalized densities of states, i.e., the number of eigenstates at each point of energy for each $D$.

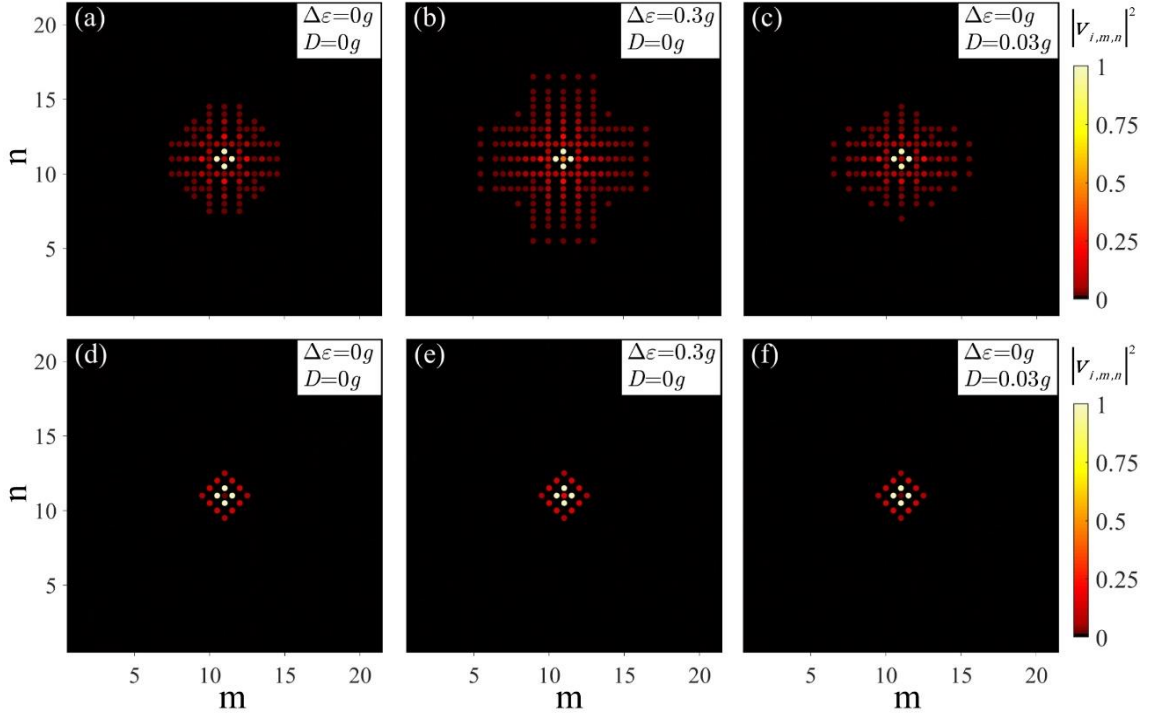

Figure 5. a)-c) Simulation results in the synthetic space without the effective magnetic flux, corresponding to parameters: a) $D=0$, $\Delta\varepsilon = 0$, b) $D=0$, $\Delta\varepsilon/g = 0.3$, and c) $D/g=0.03$, $\Delta\varepsilon = 0$. d)-f) Simulation results in the synthetic space with the effective magnetic flux, corresponding to parameters: a) $D=0$, $\Delta\varepsilon = 0$, b) $D=0$, $\Delta\varepsilon/g = 0.3$, and c) $D/g=0.03$, $\Delta\varepsilon = 0$. The color-encoded $|v_{i,m,n}|^2$ represents the energy of the corresponding photon state in the synthetic space.

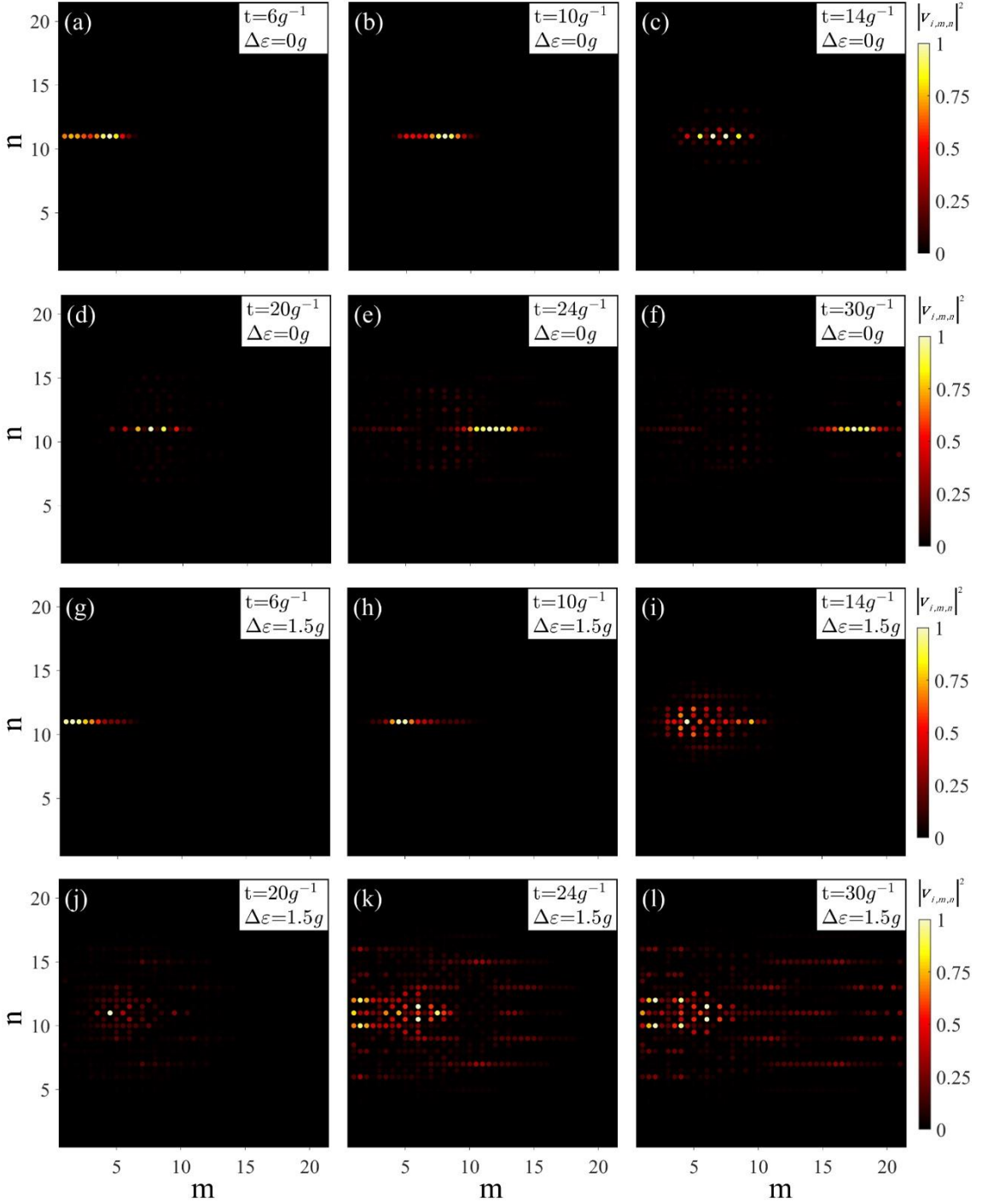

Figure 6. Simulation results in the synthetic space for the initial excitation at the site $A_{1,11}$ with the source in Equation 10 and modulations being turned on during $t \in [10, 20]g^{-1}$. Distributions of the energy of light are plotted for the excitation $\Delta\varepsilon = 0$ at different times $t = 6g^{-1}$ (a), $t = 10g^{-1}$ (b), $t = 14g^{-1}$ (c), $t = 20g^{-1}$ (d), $t = 24g^{-1}$ (e), and $t = 30g^{-1}$ (f), respectively, as well as for the excitation $\Delta\varepsilon/g = 1.5$ at different times $t = 6g^{-1}$ (g), $t = 10g^{-1}$ (h), $t = 14g^{-1}$ (i), $t = 20g^{-1}$ (j), $t = 24g^{-1}$ (k), and $t = 30g^{-1}$ (l), respectively. The color-encoded $|v_{i,m,n}|^2$ represents the energy of the corresponding photon state in the synthetic space.